\documentclass[journal]{IEEEtran}
\usepackage{cite,array,fixltx2e}

   \usepackage[dvips]{graphicx}
   \DeclareGraphicsExtensions{.eps}

\usepackage[cmex10]{amsmath}
\usepackage[caption=false,font=footnotesize]{subfig}
\begin{document}

\title{A representation of non-uniformly sampled deterministic and random signals and their reconstruction using sample values and derivatives}

\author{Nirmal~B.~Chakrabarti,~\IEEEmembership{Senior~Member,~IEEE}
\thanks{Nirmal B. Chakrabarti is with Department of Electronics and Electrical Communication Engineering, Indian Institute of Technology, Kharagpur, 721302 INDIA e-mail: nirmalbc@yahoo.com.}
}

\maketitle

\begin{abstract}
Shannon in his 1949 paper suggested the use of derivatives to increase the W*T product of the sampled signal. Use of derivatives enables improved reconstruction particularly in the case of non-uniformly sampled signals. An FM-AM representation for Lagrange/Hermite type interpolation and a reconstruction technique  are discussed. The representation using a product of a polynomial and exponential of a polynomial is extensible to two dimensions.

When the directly available information is inadequate, estimation of the signal and its derivative  based on the correlation characteristics of Gaussian filtered noise has been studied. This requires computation of incomplete normal integrals. Reduction methods for reducing multivariate normal variables include multistage partitioning, dynamic path integral and Hermite expansion for computing the probability integrals necessary for estimating the mean of the signal and its derivative at points intermediate between zero or threshold crossings.  The signals and their derivatives as measured or estimated are utilized to reconstruct the signal at a desired sampling rate.
\end{abstract}

\section{Introduction}
The commonest interpolator is a Lagrange polynomial interpolator. Widely used Whittacker-Kotelnikov-Shannon \cite{w15,k33,s49} interpolator for uniform sampling has a close relation to Lagrange interpolation.  Shannon in his 1949 paper \cite{s49} pointed out the possible application of derivatives of a signal to increase the WT product. The usefulness of derivatives in telemetry was discussed  in 1955 by Fogel \cite{f55}. The extension to non-uniform sampling was developed by Linden and Abramson \cite{l-a60} and Rawn \cite{r89}. Interestingly the theoretical framework for interpolation using a function and its derivatives was built by Hermite more than 130 years ago \cite{h78,d75}. A very large literature on interpolation and reconstruction now exists \cite{m01,m02}. Importance of timing accuracy in sampling has long been recognized (Papoulis) \cite{p66}. This requires greater attention when derivatives are used \cite{e-r-t07}.
 
The present work is concerned with a development which simplifies the computation involved in incorporating the derivatives. The classes of signals considered include natural sampling based on threshold crossing and sampling at the extrema. Methods for multivariate incomplete integration to estimate signal values from correlation characteristic of a filtered Gaussian process \cite{r45} has been studied.

Section \ref{nus} discusses the procedure for restoring local symmetry in non-uniform sampling and consequences thereof in simplifying the procedure for incorporating derivative  informations. Section \ref{es} is concerned with the framework for estimating the statistical mean of the signal and its derivatives at a desired time from a knowledge of the correlation structure. The techniques of partitioning the correlation matrix or its inverse are discussed. Attention is drawn to a path integral method in the time domain. This is based on the work of  Plackett \cite{p54,p67}. Hermite expansion \cite{k53} for computing probability integrals when direct integration proves difficult is also considered. Results and discussion are presented in Sec \ref{rd}.

\section{Interpolation for non-uniform sampling}\label{nus}
The sinc function used in WKS interpolation of uniformly sampled signals is symmetric. Chebychev polynomial interpolation uses non-uniformly spaced zeros but is symmetric about the centre. A consequence of non-uniform sampling is that the odd derivatives of the polynomial defined by zero locations are non-zero. It is useful to locally restore the even symmetry about the sampling point. The first derivative of the function

\begin{equation}
G_0(x) =\prod(1- x/a_n)*(1+x/b_n)
\end{equation}
can be removed by multiplying the product by $\exp(d1*x)$ to derive 

\begin{equation}
G_1(x) = \prod(1- x/a_n)*(1+x/b_n)* \exp(d1*x)
\end{equation}
where $a_n$ and $b_n$ give locations of zeros to the right and left respectively of the origin and $d1= \sum (1/a_n -1/b_n)$.

More generally the product function $\prod(1-x/a_n)(1+x/b_n)$ is multiplied by a symmetrizer

\begin{equation}\label{sz}
S(x) = \exp (d1*x+ d3*x^3/3 + \ldots )
\end{equation}
where $dk= (\sum 1/a_n^k- 1/b_n^k)$ for $k$ odd, i.e., to obtain $G(x)$. Thus

\begin{equation}
G(x) = S(x)*G_0(x)
\end{equation}

It is to be noted that the even derivatives are necessarily non-zero.

One  gets for the case when the first derivative ($f'(0)$) alone is to be incorporated

\begin{equation}
f(x)= f(0)*\exp(f'(0)/f(0)*x) *G(x)
\end{equation}

\begin{equation}
\mbox{Let }f(x)=A(x)*G(x)
\end{equation}
where $A(x)$ is the amplitude modulation function and $G(x)$ is the switching function or FM term for non-uniform sampling. The first few derivatives of $A(x)$ at $x=0$ can be found from the equations below if $dG/dx=0$ and $G(x)=1$ at $x=0$ as desired for any interpolator:

\begin{subequations}\label{se}
\begin{equation}
dA/dx=df/dx
\end{equation}
\begin{equation}
d^2 A/dx^2= d^2f/dx^2 - 3*dA/dx* d^2G/dx^2
\end{equation}
\begin{equation}
d^3A/dx^3= d^3f/dx^3- 6*d^2 A/dx^2* d^2G/dx^2
\end{equation}
\end{subequations}

Higher order derivatives of $A(x)$ requires a knowledge of lower order derivatives of $A(x)$ and even order derivatives of $G(x)$. A formal relation between $A(x)$ and $f(x)$ is derived from the expression for the  derivative of $f(x)/G(x)$.

For the case of zero crossing the first derivative $f'(x)$ is expressed as $f'(x)= A(x)*G(x)$ and second and higher derivatives are derived in the manner indicated. Extremum sampling is based on the amplitude and second and higher derivatives at points where the first derivative vanishes and Eqns. (\ref{se}) apply.

A useful alternative expression when $f(0)$ is not close to zero is

\begin{equation}\label{fz}
 f(x)= f(0)\exp(m(x))G(x)
\end{equation}
Letting $f_1(x)= \exp(m(x))*G(x)$, modulation function $m(x)$ is derived from the logarithmic derivative of $f_1(x)/G(x)$, where $f_1(x)= f(x)/f(0)$. The above can be stated formally as: derivatives of the exponential amplitude modulation $m(x)$ are given by the relation

\begin{equation}\label{dnm}
 \frac{d^n}{dx^n} m(x) = \frac{d^n}{dx^n} \left(\ln \left(\frac{f(x)}{f(0)}\right)\right)- \frac{d^n}{dx^n}(\ln (G(x))
\end{equation}

The second term in the R.H.S. of Eqn. (\ref{dnm}) is simply related to $dn$. $f(x)$ given by Eqn. (\ref{fz}) is seen to be a product of a polynomial and exponential of a polynomial determined by the derivatives of the signal. $G(x)$ can be raised to a desired power as in Hermite interpolation. In polynomial based generalized Hermite interpolation, the order of the polynomial for specified zero location is strictly related to the number of derivatives desired. This is relaxed in envelop FM description. It is to be noted that this operation reduces the contribution from samples distant from the point examined, thus reducing as expected the number of sample points. Imposition of local symmetry is therefore especially useful when derivatives are used.

A representation of entire functions as a product of a polynomial with specified zeros and an exponential of a polynomial is useful for nonuniform sampling. This ensures that zeros continue to occur at locations desired while the AM envelope is determined by the derivatives.

In the case of a band pass signal with in-phase and quadrature components I and Q, one may separately find the interpolated values and later combine to form a complex signal at a desired frequency.

In the two dimensional case, one may express 
\begin{equation}
f(x,y)=f(0,0)*\exp(m(x,y))*G(x,y)
\end{equation}

In separable form
\begin{equation}
G(x,y)=X(x)Y(y)
\end{equation}

Differentials in Eqn. (\ref{dnm}) are now replaced by two dimensional derivatives, i.e.,
\begin{eqnarray}
\frac{{\partial}^{r+s}}{{\partial}x^r {\partial}y^s}(m(x,y))= \frac{{\partial}^{r+s}}{{\partial}x^r {\partial}y^s}\left(\ln \frac{f(x,y)}{f(0,0)}\right)-\nonumber\\
\frac{{\partial}^{r}}{{\partial}x^r}\ln{X(x)}-\frac{{\partial}^{s}}{{\partial}y^s}\ln{Y(y)}
\end{eqnarray}

A limitation of the exponential representation is the requirement that the signal amplitude is not close to zero. This is avoidable by choice of the crossing threshold.

Symmetrizer defined by Eqn. (\ref{sz}) ensures local symmetry of the contribution of the signal at $z=0$. Approximate symmetry for a wider range restricted to narrowband applications can be established by introducing a time shift as given by Lomb \cite{l76}.

Taking the simplest case of two point interpolation, one finds that a cubic interpolation requires a knowledge of the sample value and first derivative at end points as indicated by 
$$(x-a)^2( B0+B1(x-b))+(x-b)^2.(A0+B1(x-a))$$

\begin{figure}
\centering
\includegraphics[width=3.4in]{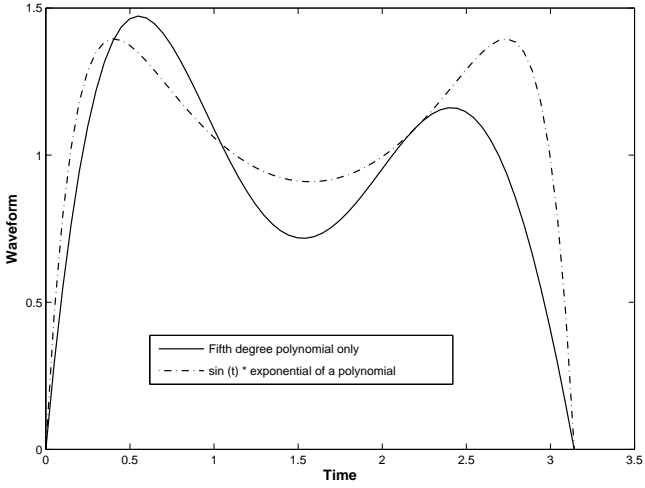}
\caption{Waveform for a polynomial only and sin($x$) multiplied by a polynomial}
\label{waveform}
\end{figure}

If one uses linear interpolation, four sample points are necessary. A general result stated in Davis \cite{d75} is: the polynomial 

\begin{equation}
p(x)\!\! =\!\! (x-a)^n\! \sum \frac{A_k}{k!}(x-b)^k +(x-b)^n\! \sum \frac{B_k}{k!}(x-a)^k
\end{equation}
with $A_k= \frac{d^k}{dx^k}[f(x)/(x-b)^n]$ and $B_k= \frac{d^k}{dx^k}[f(x)/(x-a)^n]$ satisfies the condition that the derivatives of $p(x)$ agree with the derivatives of $f(x)$ at  $a$ and $b$. For the case of nearly sinusoidal signals defined by zeros and specified slopes $s_1$ and $s_2$, one may express the function as $f(x)= \sin(x) (s_2x + (1-x)s_1)$. Use of second derivative enables one to represent functions with two maxima and a minimum as shown in Fig. \ref{waveform}, with polynomial only and $\sin(x)$ multiplied by exponential of a polynomial.

\section{Estimation of signal from correlation characteristics}\label{es}
We restrict our attention in this section to time domain signals and the symbols are chosen accordingly. The basic  assumption of the work of section \ref{es} is the presence of an underlying filter. Linear interpolation over a large number of sampling points gives rise to a sinc impulse response. When the number of sampling points is small, the use of the filter response if known enables good recovery. A stand alone two point $(0,T)$ interpolation built on the above basis may be expressed as

\begin{eqnarray}\label{yt}
x(t)= \left[x(0)m(t) +u(0)m_1(t) +w(0)m_2(t)\right]G_0(T)+ \nonumber\\
\left[x(T)m(T\!-\!t)\! +\!u(T)m_1(T\!-\!t)\! +\!w(T)m_2(T\!-\!t)\right]G(t)
\end{eqnarray}
where $x(0),x(T),u(0),u(T),w(0),w(T)$ are the values of the amplitudes and the first and second derivatives respectively at $t=0$ and $t=T$ and $G_0(T)$ and $G(t)$ are window functions which ensure that the individual sample values are not affected by the presence of other samples. $m(t)$ and $m_1(t)$ are derived from the filter impulse response. One can include higher order derivatives if these are precisely known.

The development in section \ref{nus} assumes that the values of the function and its derivatives are known. The first derivative at a sampling point may not be difficult to measure. Higher order derivatives even when measured are likely to be contaminated with noise.

In natural sampling where sampling instants are determined by crossings of  specified threshold the spacing between two successive samples may be wider than the Nyquist interval even when near equivalence of  derivatives to additional samples is assumed.

A large class of signals belongs to sampled values of a filtered Gaussian process with a specified correlation function and processes derived from the Gaussian. We assume that in the interval of interpolation a few sampled values and their derivatives are known and a few more are desired to be estimated with some measure of reliability as in the case of recovery of missing signals \cite{m01}.

\subsection{Partitioning and reduction of order of probability integral}
A $k-$variate normal distribution for a vector $\bf{x}$ with a covariance matrix, $\bf{M}$ with elements $m_{ij}$, is expressed as

\begin{equation}\label{wx1xk}
W(x_1,....,x_k) = \frac{\exp\left(-\frac{1}{2}\left({\bf{x}}^t{\bf{Ax}}\right)\right)}{(2\pi)^{k/2}\sqrt{\bf{|M|}}}
\end{equation}
where the matrix $\bf{A}$ with elements $a_{rs}$ is given by $\bf{A}=(\bf{M})^{-1}$. The elements $a_{ij}$ of $\bf{A}$ of order three quoted in many papers are for $m_{ii}=1$

\begin{eqnarray}
a_{11}=\frac{1-m_{23}^2}{{\bf{|M|}}}, a_{22}=\frac{1-m_{13}^2}{{\bf{|M|}}}, a_{33}=\frac{1-m_{12}^2}{{\bf{|M|}}},\nonumber\\
a_{12}=\frac{m_{13}m_{23}-m_{12}}{{\bf{|M|}}}, a_{13}=\frac{m_{12}m_{23}-m_{13}}{{\bf{|M|}}},\nonumber\\
a_{23}=\frac{m_{12}m_{13}-m_{23}}{{\bf{|M|}}}
\end{eqnarray}
where ${\bf{|M|}}=1-m_{12}^2-m_{13}^2-m_{23}^2+2m_{12}m_{13}m_{23}$. The partial correlation is defined as 

\begin{equation}
m_{12.3}=\frac{m_{12}-m_{13}m_{23}}{\sqrt{(1-m_{13}^2)(1-m_{23}^2)}}
\end{equation}
and other terms $m_{13.2}$ and $m_{23.1}$ are obtained by cyclical rotation.

The characteristic function corresponding to Eqn. (\ref{wx1xk}) is 

\begin{equation}
\Phi({\bf{\omega}}) = \exp(-1/2(m_{ij} {\omega_i} {\omega_j}))
\end{equation}

Three basic probability integrals which must be computed include the probability integral $P=\int W({\bf{x}}) d{\bf{x}}$; the set of means given by $m_r=\int x_k*W({\bf{x}})d{\bf{x}}$ and product moments $r_{ij} = \mbox{ave}(x_ix_j)$ = $\int x_ix_j W({\bf{x}})d{\bf{x}}$. A measure of the variance of the estimated  mean may also be required. It is known that the product moment $r_{ij}$ can be found by differentiating the probability integral $P$ with respect to $a_{ij}$. The $m_r$s are simply related to the probability integral and $\bf{M}$.

This work is concerned with threshold crossings and the effect of derivatives and side information at neighboring points. The specific problems to be studied concern estimation of (a) two point data of amplitude, the first and second derivative and (b) four point data of amplitudes and first derivatives. The amplitudes are known but the information about derivatives may be confined to their correlation behavior and the polarities. A feature which makes the algebra a bit involved is the requirement that the crossing level is ordinarily not zero. Rice \cite{r45} in his celebrated paper computed the distribution of time interval between crossings of a level from a knowledge of joint distribution of the variables $x_1(t_1),u(=dx_1/dt),x_2(t_2),v(=dx_2/dt)$ and evaluating $\int \int uv W(x_1,u,x_2,v)du dv$. An elaboration which uses the same framework is employed.

The variables are designated as $x(t_1),x(t_2)$, the first derivative $u$ and second derivative $w_2$ at $t_2,x(t_3)$, the first derivative $v$ and second derivative $w_3$ at $t_3,x(t_4)$ and $x(t)$ at a point intermediate between $t_2$ and $t_3$. Their number including the amplitude at the point of estimation for the case of two derivatives is nine. The nine-variable  density function is partitioned into two sets, one consisting of four amplitude variables $x(t_1),x(t_2),x(t_3)$ and $x(t_4)$ and other consisting of the variables including two first derivatives, two second derivatives and $x(t)$ at the time of estimation. If the slopes $u$ and $v$ are measured, the problem of estimation simplifies considerably as one is then required to find conditional distribution of $w_1,w_2$ and $x(t)$. If polarity alone of $u$ and $v$ are known, one has to consider the probability distribution $W_5(u,v,w_1,w_2,x(t))$.

\subsubsection{Partitioning for preprocessing}
The variables in the normal density function  are partitioned into two classes: (a)$x_a$, those for which the values are known and (b) $x_b$, those for which the values are not known or the polarity alone is known.

Given the correlation matrix $\bf{M(M_{11},M_{12};M_{21},M_{22})}$ and its inverse $\bf{A}(A_{11},A_{12};A_{21},A_{22})$ one can rewrite $W(\bf{x})$ as the product of $W(\bf{x_a})$ and $W(\bf{x_b/x_a})$ where 

\begin{equation}\label{wxa}
W({\bf{x_a}})=\exp(-\frac{1}{2}{\bf{x_a^t(M_{11}}^{-1}).x_a})/{(2\pi)^{m/2}\sqrt{\bf{|M_{11}|}}}
\end{equation}
and 

\begin{eqnarray}\label{wxbxa}
W({\bf{x_b/x_a}})=&\exp[-\frac{1}{2}{\bf{(x_b-R_ax_a)}}^t{\bf{A_{22}(x_b-R_ax_a)}}.\nonumber\\
&(2\pi)^{(n-m)/2}\sqrt{\bf{|A_{22}|}}]
\end{eqnarray}
where $\bf{R_a= M_{21} (M_{11}^{-1})}$ and ${\bf{|A_{22}|}}={\bf{|M_{11}|}}/{\bf{|M|}}$. Eqn. (\ref{wxbxa}) shows that integration regime for $x_b$ is modified due to the shifts caused by $x_a$. The informations contained in $W\bf{(x_a)}$ influence in two distinct ways. The first is to increase the order of the correlation matrix and the second is to introduce effective signals $\bf{S}$ represented by $\bf{M_{21}.(M_{11}^{-1})x_a}$. One can rewrite Eqn. (\ref{wxbxa}) as

\begin{equation}\label{wxbxa3}
W({\bf{x_b/x_a}})= C.\exp(-1/2.({\bf{x_b-S}})^t {\bf{A_{22}(x_b-S)}})
\end{equation}
This has an equivalence in chf which is useful when one utilizes Hermite expansion.

A simple useful example of partitioning is computation of time interval between upward crossing at $x=h$ followed by a downward crossing at threshold $k$ using the method due to (Rice) for finding crossing time in the two point case. The joint distribution of two space variables with correlation $m$ and slopes thereat may be written in partitioned form as

$$Wa(x_1,x_2,m).W(u_1,u_2/(x_1,x_2))$$
where $W_a(x_1,x_2)= \exp\left(-\frac{1}{2}({\bf{x}}^t {\bf{M_{11}^{-1}x}})\right)/(2\pi.|{\bf{M_{11}}}|)$ and $W_b({\bf{u,S}})= \exp(- {\bf{(u-S)}}^t {\bf{A_{22}.(u-S)}}).\bf{|A_{22}|}/2\pi$ where ${\bf{A_{22}}}= ({\bf{M_{22}-M_{21} (M_{11})^{-1}.M_{12}}})^{-1}$

The signals $s_3$ and $s_4$ are:

$$s_3=\frac{(m_{13}-m_{23}m_{12})a_1+(m_{23}-m_{13}m_{12})a_2}{(1-m_{12}^2)};$$
$$s_4=\frac{(m_{14}-m_{24}m_{12})a_1+(m_{24}-m_{14}m_{12})a_2}{(1-m_{12}^2)}$$

The two dimensional integral $\int uvW(u,v) dudv$ can be found by integrating $B(h,k;\rho)$. When $h$ and $k$ are both positive, one notices that in the region of interest for crossing($u>0,v<0$), $s_3$ is negative thus cutting off low positive velocity components. Similar remark applies for the negative velocity. The end result is that the integration interval is reduced and therefore the time window becomes smaller. The probability of crossing for positive threshold is also small because of dependence on threshold.

When $h=k=0$, one has the well known formula

$$R(\tau)\!=\!\mbox{ave}(uv)(u\!>\!0,v\!<\!0)\!= \!\frac{1}{2\pi}\!\!\left(\!\!r_1\!\!\left(\frac{\pi}{2}\!+\!\sin^{-1}\!r_1\!\!\right)\!\!+\!\sqrt{\!1\!-\!r_1^2}\!\right)$$
where $r_1$ is the correlation between $u$ and $v$. The effect of neighboring sampling points can be studied by enlarging the sub-matrix $\bf{M_{11}}$ and finding equivalent signals controlling the timing window.

The conditional chf corresponding to (\ref{wxbxa}) may be written as

\begin{equation}\label{pwbxa}
\Phi(\omega_b/x_a)= \Phi(\omega_b) \exp(j\omega_b.S)
\end{equation}
where $\omega_b$ is the transform variable associated with $\bf{x_b}$.

The conditional probability density function $W(x_b/x_a)$ of dimension five for the two point interpolation problem mentioned above may be written as

\begin{equation}\label{wxbxa2}
W(x_b/x_a)\!=\!W_5(u_1,w_1,u_2,w_2,x(t)/x_1,x_2,x_3,x_4) 
\end{equation}
$W_5$ yields the conditional expected value of $x(t)$ on necessary integration.

More generally let $L_0,K_0$, and $M_0$ be the number of such amplitude, slope and second derivative values that are known by measurement and let $L$ equal number of unknown amplitude variables which constrain the polarity, $K$ equal number of slope variables for which only the polarity is specified and $M$ that of number of variables for second derivatives for which the polarity alone is specified and $N$ be the number of variables for which average value has to be found.

The dimension of the probability integral is then 

\begin{equation*}\label{kL}
k=L_0+L+K_0+K+M_0+M
\end{equation*}
The $k-$dimensional density function can be reduced by utilizing the knowledge of  $L_0+K_0+M_0$ variables to obtain a density function of  dimension $n=L+K+M$. The conditional $n-$dimensional probability density function yields the expected amplitude and derivative values by appropriate integration. Following the first partitioning to introduce the known variables, further partitioning of $W(x_b)$ has to be carried out.

\subsection{Statistical mean}
A useful general result is that the statistical means of  n-variables may be expressed in terms of $n$ probability integrals of $(n-1)$ variables.

Expressing $W(\bf{x})$ as

\begin{equation}
W({\bf{x}})= C.\exp\left(-\frac{1}{2}\bf{Q(x)}\right)
\end{equation}
where ${\bf{Q(x)}}=\sum a_{rs} x_s x_r$ and $C=\frac{1}{(2\pi)^{n/2}\sqrt{|\bf{M}|}}$, one gets 

\begin{equation}\label{dwdxr}
\frac{{\partial}W}{{\partial}x_r}= -\left(\sum a_{rs} x_s\right)W
\end{equation}

Integration of equation (\ref{dwdxr}) yields $\sum a_{rs}m_s$, where $m_s$ is the statistical mean of the $s-$th variable, as $P_r$. $P_r$ is the integral of $n-1$ variables with $x_r$ replaced by its value at the lower limit.

\begin{subequations}\label{pr}
\begin{eqnarray}
P_r=C\int-\frac{{\partial}Q}{{\partial}x_r} \exp\left(-\frac{1}{2}Q\right) dx_1 dx_2\ldots dx_n\nonumber\\
 = C\int \frac{{\partial}\exp(-Q)}{{\partial}x_r} dx
\end{eqnarray}
\begin{equation}\label{q1}
\sum(a_{r1} xm_1\! + a_{r2} xm_2\! +..)\!=C\!\!\int \!\!\frac{\exp(-\frac{1}{2}Q_r)}{(2\pi)^{n/2}\sqrt{|M|}}dx'
\end{equation}
\end{subequations}
where $Q_r$ is obtained by replacing $x_r$ in $Q$ by $a_r$ and integration of RHS is carried out over $(n-1)$ variables.

The LHS of (\ref{q1}) is a weighted sum of means.

The set of simultaneous equations for the vector ${\bf{X_m}}[xm_1,\ldots,xm_n]$

\begin{equation}\label{axm}
{\bf{A.X_m}}=(P_1,P_2,P_3, \ldots, P_n)
\end{equation}
has the solution 

\begin{equation}\label{xm}
\bf{X_m= M. (P)}
\end{equation}
where $\bf{P}$ is the vector $(P_1,P_2,P_3, \ldots, P_n)$. Derivation of Eqn. (\ref{axm}) assumes that if $Q(x)$ contains a linear term, the variables are transformed to remove it. When the effect of the signals $\bf{S}$ as in Eqn. (\ref{wxbxa3}) is included, Eqn. (\ref{xm}) becomes

\begin{equation}\label{x_m}
\bf{X_m= M. (P)+S}
\end{equation}
One can associate a signal flow diagram with the above equation. It is necessary to note that $\bf{M}$ is the correlation matrix of the partitioned variables.

The first moment for a correlated pair of variables provides the simplest example and is given for the threshold at zero by

$$\mbox{ave}(x_1)=\mbox{ave}(x_2)=\frac{1}{2\sqrt{2\pi}}(1+\rho)$$
If threshold are at $h$ and $k$,

\begin{subequations}\label{x1x2}
\begin{equation}
\mbox{ave}(x_1) = \frac{P_1+\rho P_2}{2\sqrt{2\pi}}\mbox{ and }\mbox{ave}x_2\frac{P_2+\rho P_1}{2\sqrt{2\pi}}
\end{equation}
\begin{equation}
\mbox{where }P_1 = e^{-h^2/2} \mbox{erfc}\left(\frac{k-h\rho}{\sqrt{2(1-\rho^2}}\right)\mbox{ and}
\end{equation}
\begin{equation}
P_2 = e^{-k^2/2} \mbox{erfc}\left(\frac{h-k\rho}{\sqrt{2(1-\rho^2}}\right)
\end{equation}
\end{subequations}

It is well known that a two-dimensional probability integral can be reduced to a single integral.

\begin{eqnarray}\label{int-2}
\int_h^\alpha \int_k^\alpha W(x,y,\rho) dx dy =\nonumber\\
\int_0^\rho \exp\left[- \frac{h^2-2h.k.\rho + k^2}{2(1-\rho^2)}\right]. d\rho/(2\pi\sqrt{(1-\rho^2)}
\end{eqnarray}

It is convenient to use Owen's \cite{o56} procedure for computing 2D normal integrals using $T-$ functions, where $T$ is defined as
\begin{equation}\label{tha}
T(h,a)= \frac{1}{2\pi}\int_0^a \frac{\exp[-\frac{1}{2}h^2(1+x^2)]}{1+x^2}
\end{equation}

The result is 

\begin{eqnarray}\label{bhkrho}
B(h,k;\rho)= T\left(h,\frac{k}{h}\right) + T\left(k,\frac{h}{k}\right) -\nonumber\\
T\left(h,\frac{k-\rho h}{h\sqrt{1-\rho^2}}\right) -T\left(k,\frac{h-\rho k}{k\sqrt{1-\rho^2}}\right) + G(h)G(k)
\end{eqnarray}
where $G$ is one dimensional Gaussian integral. 

For the case of three variables,

\begin{equation}
\left[\!\!
\begin{array}{c}
\mbox{ave}(x_1)\\
\mbox{ave}(x_2)\\
\mbox{ave}(x_3)
\end{array}
\!\!\right] = 
\left[\!\!
\begin{array}{c c c}
m_{11} & m_{12} & m_{13}\\
m_{21} & m_{22} & m_{23}\\
m_{31}&  m_{32} & m_{33}
\end{array}
\!\!\right] 
\left[\!\!
\begin{array}{c}
B(a_1,x_2,x_3)\\
B(a_2,x_1,x_3)\\
B(a_3,x_1,x_2) 
\end{array}
\!\!\right] 
\end{equation}
where $B(a_i,x_j,x_k)$ signifies a two dimensional integral with $x_j$ replaced by $a_j$ in $Q(x)$. The result can be expressed in terms of partial correlations. When $a_1=a_2=a_3=0$, one gets

\begin{eqnarray}
\mbox{ave}(x_1)= \frac{1}{2\pi^{3/2}}\left(\frac{\pi}{2}+ \sin^{-1}(m_{23.1})\right)+ \nonumber\\
m_{12}\left(\frac{\pi}{2}+\sin^{-1}(m_{12.3})\right)+ m_{13}\left(\frac{\pi}{2}+\sin^{-1}(m_{13.2})\right)
\end{eqnarray}
The first is a self term and the other two are mutual terms representing contributions from other two equivalent sources. For the case of four sources, four three dimensional integrals $P(a_1,x_2,x_3,x_4)$, $P(x_1,a_2,x_3,x_4)$, $P(x_1,x_2,a_3,x_4)$, $P(x_1,x_2,x_3,a_4)$ multiplied by the coupling terms generate the means. The integrals in the five dimensional case may be written as $P(a_i,x_j,x_k,x_l,x_m)$ which arise from integrating $C.\exp(-1/2.Q(a_i,x_j,x_k,x_l,x_m)$ with $x_i$ replaced by $a_i$ in $Q(x)$.

\subsection{Relating a probability integral to a set of integrals of lower dimension}
Reduction of probability integral and integral for finding mean to a set of integrals of lower order is considered in this section.
The probability integral, $P_n(a_1,a_1,a_3..a_n)$ is computed from

$$P_n(a_1,a_2,a_3..a_n) = \!\! \int \!\! \int \!\! \int \!\! \int \! W(x_1,x_2,x_3..) dx_1 dx_2 dx_3 ..dx_n$$
where $a_1$ to $a_n$ are the lower limits while the other limit is infinity.

The general method due to Plackett for reduction of order may be stated briefly as follows.

\begin{equation}\label{partialp}
\frac{\partial P}{\partial (r_{12})}= \int \int W(a_1,a_2,x_3,\ldots x_n.) dx_3 \ldots dx_n
\end{equation}

This can be derived directly or from the partial differential equation (Plackett 1954)

\begin{equation}\label{partialw}
\frac{\partial W}{\partial (r_{ij})}=\frac{\partial^2 W}{\partial {x_i} \partial {x_j}}
\end{equation}

Use of (\ref{partialw}) yields

\begin{equation}\label{dw}
\frac{\partial W}{\partial (r_{12})}= W_2(h,k,r_{12})* W_{n-2}(x_d,A_{22},s_1,s_2)
\end{equation}

A method of considerable power is based on conditional probabilities (Steck \cite{s58}) which works from low order to a higher order. A three dimensional integral can be expressed as a sum of three two dimensional integrals and four-dimensional probability integrals as sum of four three dimensional integrals. It is well known that computation of such integrals for $n>3$  poses difficulties. Several methods are available for reduction of the probability integrals principally through partitioning for the variables into smaller groups in the manner stated earlier.

Path integral: Plackett devised a line integral method for finding the probability integral for a point P defined by the correlation matrix M once the value at another point K is known. Symbolically for $i \neq j$

\begin{equation}
m_{ij}=(1-t)m_{ij}(K)+t m_{ij}(P)
\end{equation}

A modification proposed by Pawula is to multiply off-diagonal terms by $t$ in the correlation matrix.

Dynamical path integral: In physical problems concerning filtered Gaussian noise where the correlation function and its derivatives can be derived, a direct approach is to form the time derivative and use the relation

\begin{subequations}\label{dpi}
\begin{equation}
\frac{\partial W}{\partial t} = \sum \frac{\partial m_{ij}}{\partial t} * \frac{\partial W}{\partial m_{ij}}
\end{equation}
\mbox{to derive}
\begin{equation}
\frac{\partial P(n)}{\partial t} = \sum \frac{\partial m_{ij}}{\partial t}*P(n-2)
\end{equation}
\end{subequations}

Partitioning variables as in (\ref{wxbxa}), this can be written as 

\begin{equation}
\frac{\partial P}{\partial t} =\sum \dot{m_{ij}} \int W(a_1,a_2,{\bf{A_{kl}}},x_k,x_l)dx_k dx_l
\end{equation}
where the matrix $\bf{A_{kl}}$ corresponds to $\bf{A_{22}}$ in Eqn. (\ref{wxbxa}).

For $n=3$ one gets 

\begin{eqnarray}\label{dopdot}
\frac{\partial P}{\partial t} =\dot{m_{12}} \int W(a_1,a_2,x_3)dx_3 +\nonumber\\
\dot{m_{13}} \int W(a_1,a_3,x_2)dx_2 +\dot{m_{23}} \int W(a_2,a_3,x_1)dx_1
\end{eqnarray}

The (1,2) component of (\ref{dopdot}) may be written as 

\begin{eqnarray}\label{dop12dot}
\frac{\partial P_{12}}{\partial t} =&\frac{\dot{m_{12}}}{(2\pi)^{3/2}\sqrt{1-m_{12}^2}}.\nonumber\\
&\exp\left(-\frac{1}{2}\frac{a_1^2-2m_{12}a_1a_2+a_2^2}{1-m_{12}^2}\right)\nonumber\\
&\sqrt{a_{33}} \int \exp\left(-\frac{(x_3-S_1)^2a_{33}}{2}\right) dx_3 
\end{eqnarray}
where $S_1=\frac{((m_{13}-m_{23}m_{12})a_1 + (m_{23}-m_{13}m_{12}))a_2}{(1-m_{12}^2)}$.

The integral can be expressed as an error function with argument $S_1$.

If $a_1=a_2=a_3=0$, one has a simple result

\begin{eqnarray}
\frac{\partial P}{\partial t} =\frac{1}{4\pi}\left(\frac{\dot{m_{12}}}{\sqrt{1-m_{12}^2}} +\right.\nonumber\\
\left.\frac{\dot{m_{13}}}{\sqrt{1-m_{13}^2}}  +\frac{\dot{m_{23}}}{\sqrt{1-m_{23}^2}}\right) 
\end{eqnarray}
This can be readily integrated to derive the well known expression $P_3=\frac{\sin^{-1}(m_{12})+ \sin^{-1}(m_{13})+ \sin^{-1}(m_{23})+ \pi/2}{4\pi}$. The advantage of the approach lies in providing intermediate values of the integrals from small or no correlation to the present correlation values. This may be considered to be a backward evolution starting with large time separation when correlations are negligible. A use of the dynamical path integral is to find $\frac{\partial P}{\partial a_{rs}}$ by computing derivatives of $P$ and $a_{rs}$ and then dividing the results. When the time derivative is not known, Plackett's method or Pawula's version may be used. A variation is to use time values where some correlations are negligible and therefore integration is easy to carry out.

\subsection{Chf based computation - Hermite expansion}\label{chf}
It is well known that multivariable expansion of the characteristic function and its inversion is useful for evaluating probability integrals and statistical means. The starting step in this case too is partitioning and  including the effect of known signals on the uncertain ones. The  transform space $\omega$ is segmented into $\omega_a$ and $\omega_b$ and transform $\Phi$ into $\Phi_a(\omega_a), \Phi_b(\omega_b)\mbox{ and }\Phi_{ab}(\omega_a,\omega_b)$, i.e.,

\begin{equation}
\mbox{chf}(\omega)= \Phi_a(\omega_a). \Phi_b(\omega_b). \Phi_{ab}(\omega_a,\omega_b)
\end{equation}

Integration over $\omega_a$ gives
\begin{equation}
\Phi(\omega_b/x_a) =\int \Phi(\omega) \exp(j\omega_a.x_a) d\omega_a
\end{equation}

This is equivalent to Eqn (\ref{pwbxa}). Hermite expansion is then carried out over $\omega_b$. We consider the specific problem of computation of four variable integral. Let a correlation matrix have non-diagonal elements defined as $m_{12}$, $m_{13}$, $m_{14}$, $m_{23}$, $m_{24}$, $m_{34}$.

The expansion of the chf contains typical terms 
$$\frac{m_{12}^p.m_{13}^q.m_{14}^r.m_{23}^s.m_{24}^t.m_{34}^u}{(p!)(q!)(r!)(s!)(t!)(u!)}* \omega_1^{m_1}\omega_2^{m_2}.\omega_3^{m_3}.\omega_4^{m_4}$$
where $m_1= p+q+r; m_2=p+s+t; m_3=q+s+u; m_4= r+t+u$. Noting that multiplication by a power of $w$ is equivalent to differentiation of that order of the individual Gaussian variable, one can find the probability integrals and the means.

Application of Hermite expansion is limited to three variables with three off-diagonal terms. Direct four variate expansion is expensive. The number of terms for the triple product is as large as 216. An alternative using bivariate integrals and their derivatives is useful.

chf can  be partitioned as

$$\Phi= \Phi_a \Phi_b. \Phi_{ab}$$
where $\Phi_a$ and $\Phi_b$ are the chfs of the variables $xa$ and $xb$ and $\Phi_{ab}$ represents the mutual correlation terms. A four variable chf may be written as a product of  chfs  $\Phi_a$ and $\Phi_b$ of two pairs of correlated variables and a term $\Phi_{ab}$ representing mutual correlation. The density function is then a product of the density functions $W_a(x_1,x_2,m_{12})$ and $W_b(x_3,x_4,m_{34})$ operated upon by terms resulting from expansion of $\Phi_{ab}$. $\Phi_{ab}$ may be written as

\begin{equation}
\Phi_{ab} = \exp (- ( m_{13}\omega_1\omega_2+m_{14}\omega_1\omega_4+m_{23}\omega_2\omega_3+m_{24}\omega_2\omega_4))
\end{equation}

The expansion differs from usual single variable expansions as it groups a pair of variables and simplifies computation of probability integrals. As an example the probability integral and the mean for one sided variables for four variates can be expressed as power series in $m_{12},m_{13},m_{14},m_{23},m_{34}$ and term by term integration may then be carried out. One may also divide the variables into two groups and the mutual correlation chf $\Phi_{ab}$ may be expanded as

\begin{eqnarray}\label{sums}
\Phi_{ab} = \sum \sum \sum \sum \frac{m_{13}^p.m_{14}^q.m_{23}^r.m_{24}^s}{(p!)(q!)(r!)(s!)}\nonumber \\
* \omega_1^{m_1}\omega_2^{m_2}.\omega_3^{m_3}.\omega_4^{m_4}
\end{eqnarray}
where $m_1=p+q,m_2=r+s,m_3= p+r$ and $m_4= q+s$.

The probability density function can now be expressed as:

\begin{eqnarray}\label{fa}
F_a( D1^{m_1},D2^{m_2})*W_2(x_1,x_2,m_{12})* \nonumber\\
F_b (D3^{m_3},D4^{m_4})*W_2(x_3,x_4,m_{34})
\end{eqnarray}
where $D1=d/dx_1,D2=d/dx_2,D3=d/dx_3,D4=d/dx_4$. $F_a$ and $F_b$ represent sums 
resulting from inversion of (\ref{sums}) and $W_2$ are two dimensional integrals.

\subsection{Comment}
The results of Sec. \ref{es} can be extended for estimating signals for some non-Gaussian processes describable as Gaussian process with random parameters. A special class is sub-Gaussian symmetric alpha-stable process. A stable random vector may be expressed as $X=\sqrt(A)*G$ where $G$ is a  Gaussian process. If A has a Laplace transform of the form $\exp(-(s) \alpha/2)$ where $s$ is the transform variable, the chf of the sub-Gaussian process is given by

\begin{equation}
\mbox{chf}\alpha(\omega)=\exp(- (Qg(\omega)^\alpha/2)\nonumber
\end{equation}
where $Qg$ is the exponent of the chf of a Gaussian process. 

A  non-Gaussian process  generated with random scaling of  a Gaussian process is described in Grigoriu \cite{g95}. When the scale parameter of variance has an inverted gamma distribution, the unconditional density function on averaging over the scale parameter becomes

$$p(x)\!= \!\frac{\Gamma((m+n)/2)}{\Gamma(n/2)(n\pi\beta^2)^{m/2}\sqrt{|{\bf{M}}|}}\!\left(\!\!1\!+ \frac{1}{n\beta^2}{\bf{x^t M^{-1} x}}\!\!\right)^{\!\!-(m+n)/2}$$
For n=1,this becomes a multidimensional Cauchy distribution.

An approach applicable for filtered Poisson process is a sum of Gaussians with distinct covariances. It is known that some combinations of amplitude distribution of impulses and the filter characteristics result in nearly Gaussian distribution. The conditional distribution for a given number of impulses in an interval within the filter time window therefore generates a Gaussian sum.

\section{Results and discussion}\label{rd}
It is known that there is an equivalent Nyquist rate for the case of non-uniform sampling. The condition for reconstruction is that the sampling interval rate lies within a specified range of the Nyquist interval. This restriction is relaxed when derivatives are available. The Hermite interpolation reconstruction procedure requires that the order of the polynomial be doubled for a first derivative compared to the order for signal amplitude only. A legitimate procedure is to find the inter-sampling interval and compute the necessary number of derivatives. If this alternative is not implementable, one has the option of estimating values of the signal using characteristics of signal.

The signals are therefore a mixture of deterministic and partially known components. The number of components belonging to the second category is restricted by the requirement of  evaluation of probability integrals. As noted in Sec. \ref{es}, it is not difficult to compute the statistical mean of five variables based on computation of four probability integrals. The two point problems with known terminal amplitudes are easily solvable. These include finding (a) mean of slopes at two points and amplitude at an intermediate point; (b) two second derivatives and an amplitude if slopes are known, (c) two first and second derivatives each at terminal points and an amplitude. For the four point case, the amplitudes at four points are known. If the slopes are also known, one can estimate four second derivatives and an amplitude at an intermediate point. 

The procedure can be stated as follows. The correlation matrix and the density function and chf are first partitioned for reduction to partially known variables and those to be estimated. Probability integrals necessary for finding the statistical means are then computed. This step requires partitioning and reduction to simple one dimensional integrals. 

The known values of the samples and their derivatives are then combined with those estimated for an FM-AM representation. This is then converted to uniformly sampled representation.

The filter chosen for computation has a correlation characteristics given by $r(t) = \exp(-at^2).sinc(t)$. This is used to form the correlation matrix of the amplitudes, and the first and second derivatives in the manner discussed by Rice(BSTJ,1945). The effective signal value is then computed for a specified set of sampling points adjoining the region of estimation. The parameter $a$ permits one to cover the behavior from monotonic to pure sinc filter. In natural sampling the sampling event occurs when a threshold is crossed. The time interval between two events depends on a combination of filter characteristics and local energy.

\begin{figure*}
\begin{minipage}[b]{0.32\linewidth}
\centering
\includegraphics[width=2.25in]{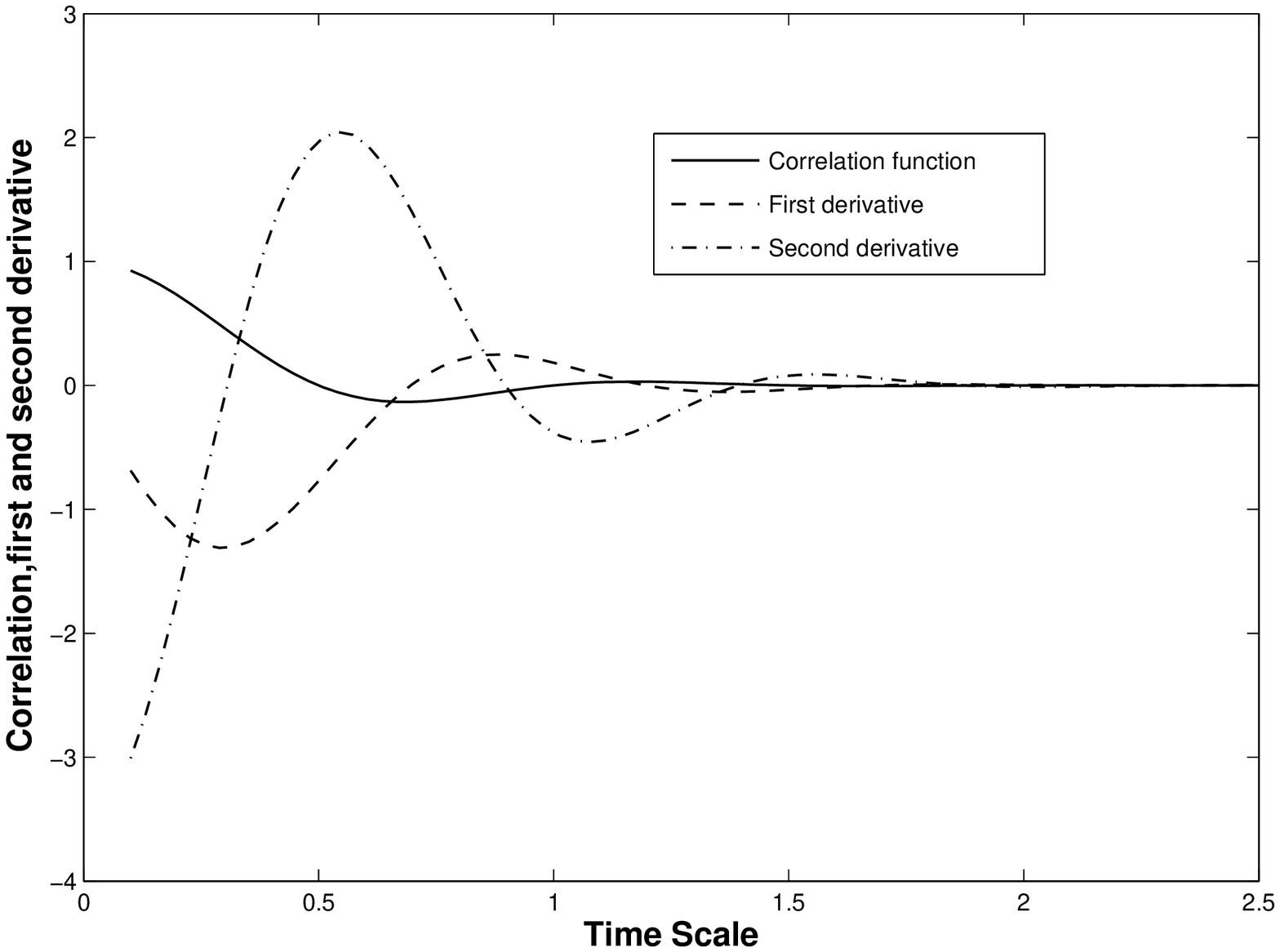}
\caption{Correlation characteristics given by $r(t) = \exp(-at^2).sinc(t)$ and the first two derivatives}
\label{corr.char}
\end{minipage}
\hspace{0.1cm}
\begin{minipage}[b]{0.32\linewidth}
\centering
\includegraphics[width=2.25in]{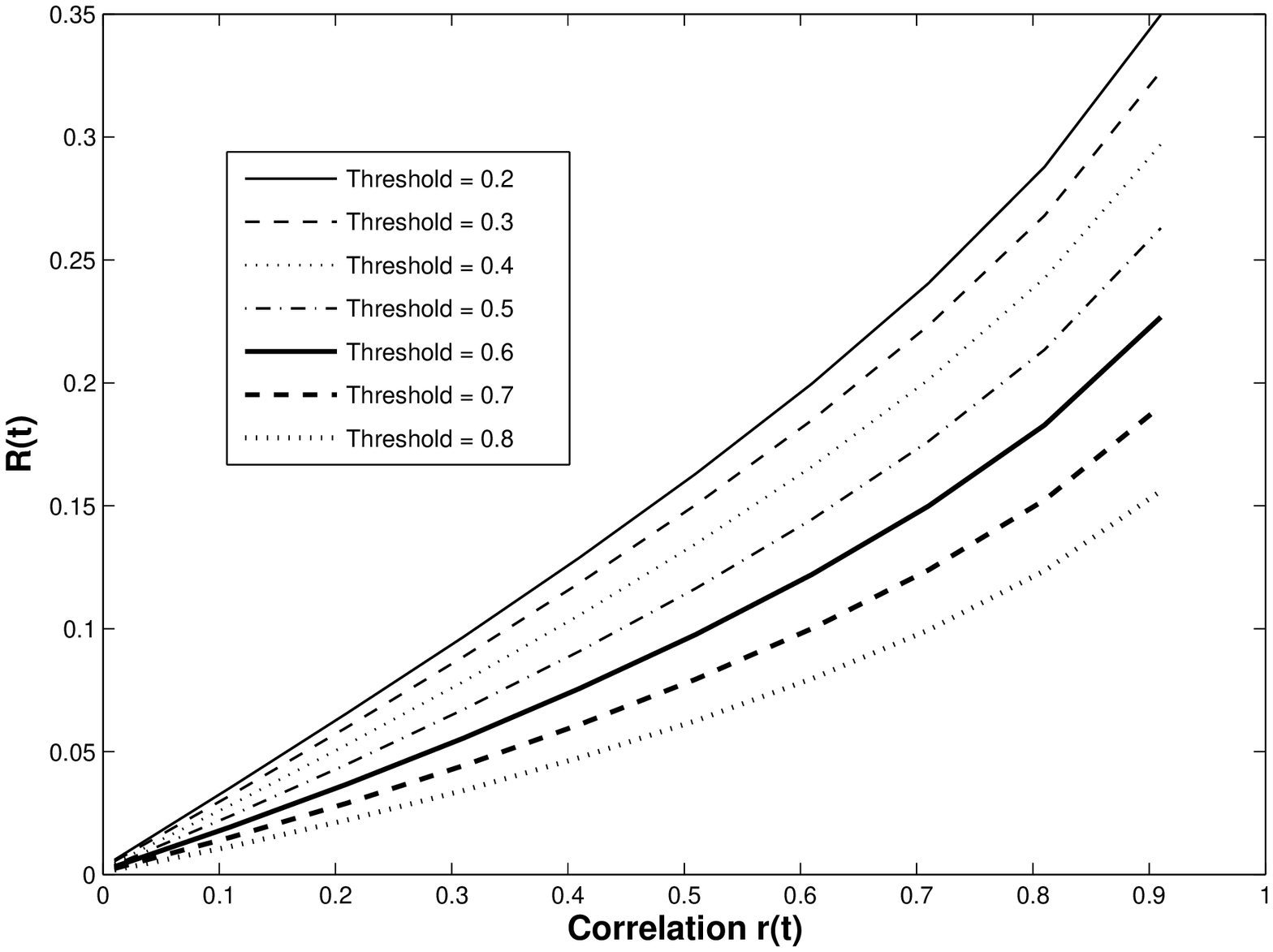}
\caption{Variation of $u-v$ correlation $R(\tau)$ with thresholds $h$ and $k$}
\label{Rt}
\end{minipage}
\hspace{0.1cm}
\begin{minipage}[b]{0.32\linewidth}
\centering
\includegraphics[width=2.25in]{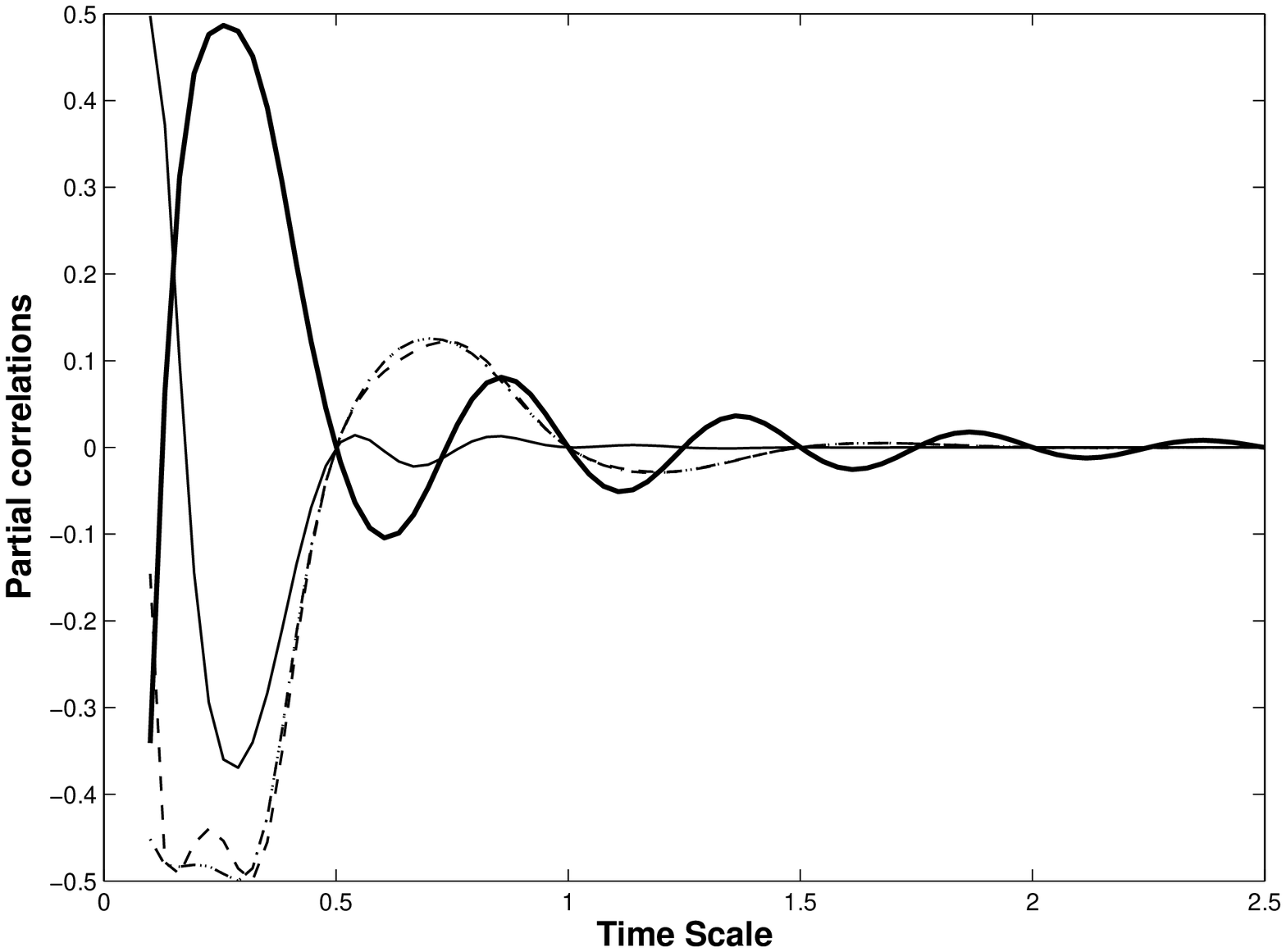}
\caption{Partial correlations}
\label{part.corr}
\end{minipage}
\end{figure*}

Figure \ref{corr.char} shows $r(t), dr/dt$ and $d^2r/dt^2$ for a particular choice of $a=0.25$. The zero of $r(t)$ occurs at that of $sinc(t)$. The event pair of an upward crossing followed by a downward crossing is related to the correlation between the slopes at time interval $t_{12}$. The polarity and amplitude of second derivative indicate the time window permitted for such events to take place. The value of the threshold has an immediate control on the actual time window.

If the thresholds $h$ and $k$ lie on the same side of zero, $R(t)$ is compressed as shown by Fig. \ref{Rt} which assumes $h=k$ and local energy parameter is unity. The control of $h$ and $k$ occurs through the medium of effective signal mentioned in Sec. \ref{es}. The monotonic relationship between $R(t)$ and $r(t)$ remains unchanged though.

\begin{figure*}
\begin{minipage}[b]{0.32\linewidth}
\centering
\includegraphics[width=2.25in]{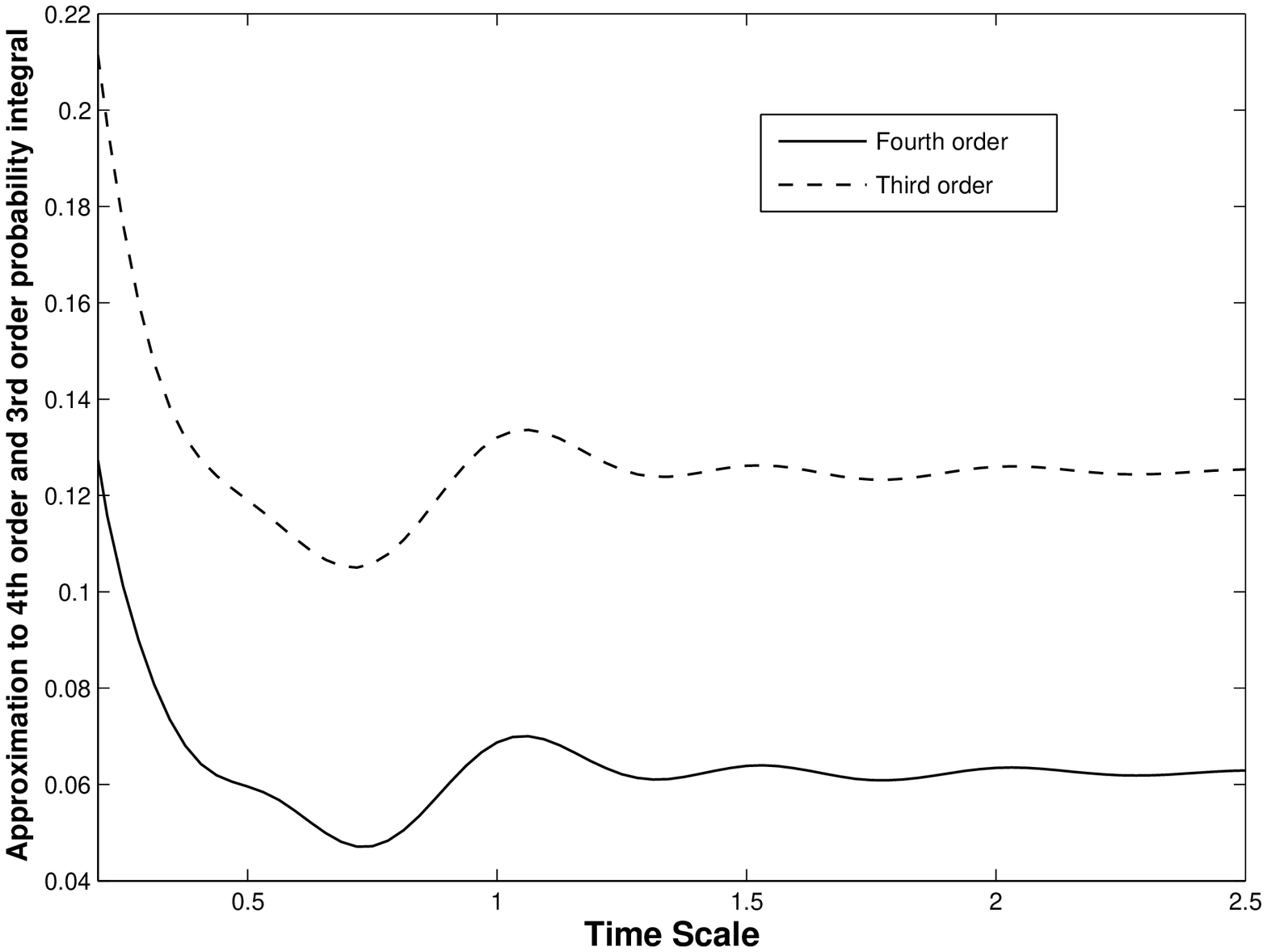}
\caption{Approximations to fourth order and third order probability integrals}
\label{approx.43}
\end{minipage}
\hspace{0.1cm}
\begin{minipage}[b]{0.32\linewidth}
\centering
\includegraphics[width=2.25in]{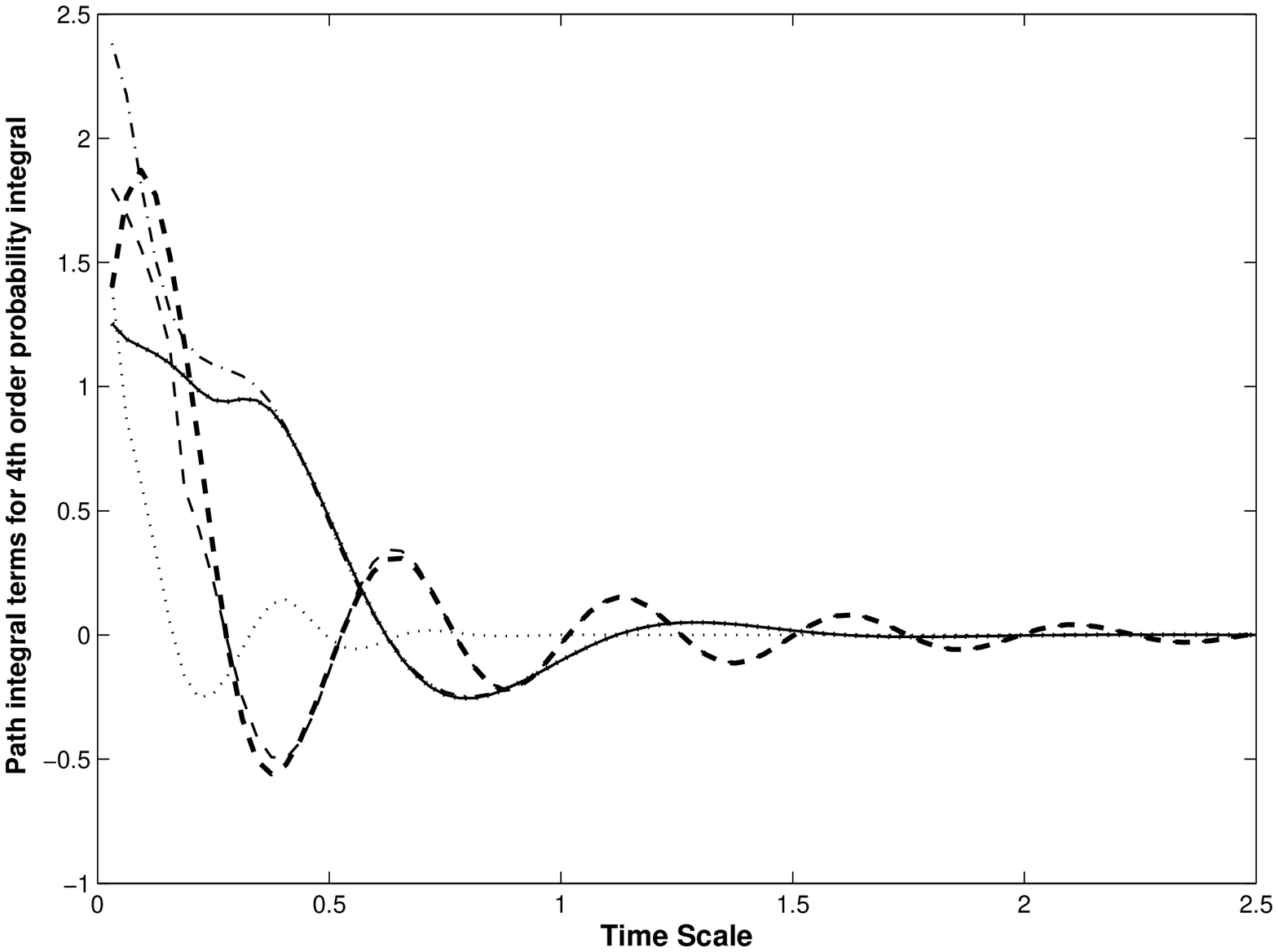}
\caption{Path integral terms of the fourth order probability integral}
\label{ind.pit}
\end{minipage}
\hspace{0.1cm}
\begin{minipage}[b]{0.32\linewidth}
\centering
\includegraphics[width=2.25in]{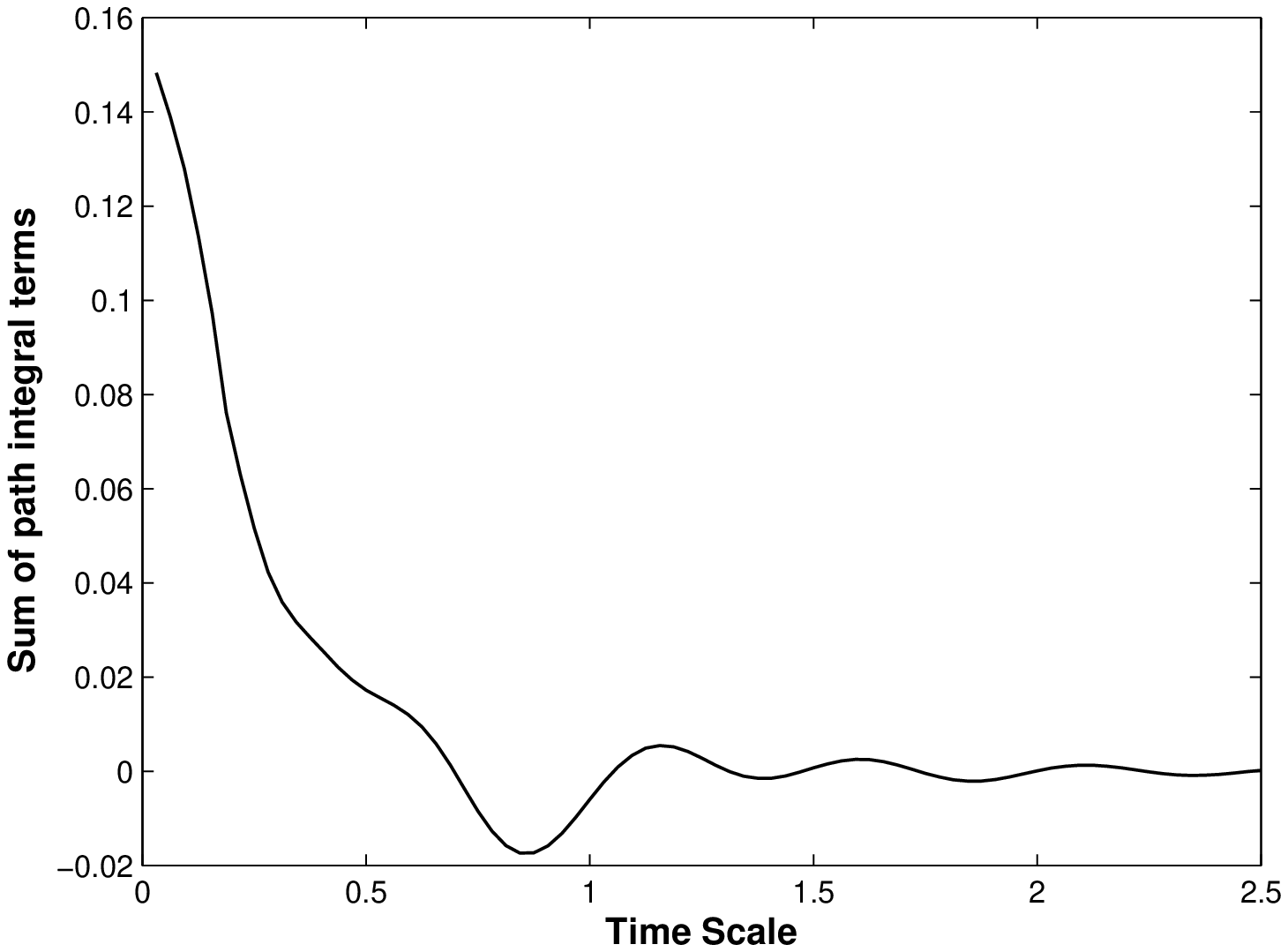}
\caption{Sum of path integral terms and time integral}
\label{sum.pit}
\end{minipage}
\end{figure*}

It is important for computations based on dynamic time integral to find how the value of the determinants changes with time scale. Once the correlation matrix is partitioned, the correlation behavior becomes conditional. The partial correlations for a four variate case are plotted in Fig. \ref{part.corr}. If the partial correlations have low values, the computational complexity can be reduced. Figure \ref{approx.43} shows the approximations to fourth order and third order probability integrals. Figure \ref{ind.pit} shows the individual path integral terms of the fourth order probability integral and their sum is shown in Fig. \ref{sum.pit}. A final time integral upto the time scale of interest yields the desired result.

A knowledge of the correlation characteristics and local energy is the basis of signal estimation discussed in sec 3; this does not appear to be simple when the time window is very short. A straight forward way is to use the data about the amplitude and the derivatives near the estimation region to interpolate and resample at uniform high rate and then find local correlation features. Teager \cite{k93} energy operator which uses the signal and its first two derivatives offers another means. The operator is expressed as 

\begin{subequations}
\begin{equation}
TK= (dx/dt)^2- x(t) d^2x/dt^2 
\end{equation}
\mbox{In discrete form it is written as} 
\begin{equation}
TK=x(n)^2-x(n-1)*x(n+1) 
\end{equation}
\end{subequations}

where $x(n)$ is the $n$th sample. This can be interpreted as $r(0)*(1-m(t))$ where $m(t)$ is normalized correlation for small time $t$ between $x(n-1)$ and $x(n+1)$. A sum of the operator values over a large number of samples may be used to indicate the second moment of the spectral density. 

The number of crossings at different levels and the time interval between crossings can be utilized for deriving informations about the energy and correlation. It is known that the number of  zero-crossings $N0$ is related to $d^2r(t)/dt^2$ at $t=0$ and equivalently to the second moment of the spectrum. The crossing rate at a height $h$ is obtained by multiplying $N0$ by $\exp(-h^2/2\sigma^2)$. A comparison of the crossing rates at different heights will therefore give a measure of $\sigma$. The timing distribution at a zero crossing gives a measure of  $d^2r(t)/dt^2$ and time between consecutive crossings at $h$ and $k$ furnishes a measure of effective signals and therefore $dr/dt$. 

Statistical averages of amplitude or derivative contain a parameter representing corresponding correlation. When the agreement between the actual measurements and the estimates is satisfactory, one can extract the correlation parameter for the time separations involved. 

Concluding remarks; Symmerization of Lagrangian interpolation function is seen to be useful in incorporation of derivatives and leads directly to an envelope-FM representation. When signal derivatives are not precisely known, the values of  signal and its derivative at a point intermediate between sampling points can be found from a knowledge of the correlation characteristic. Two stage partitioning to reduce dimensionality of the probability integrals involved, once to make use of knowledge of signals and the derivatives and then to compute the conditional probability integrals has been shown to be an essential tool. Dynamical path integral method commends itself when values for a continuously scaled time instants are desired. A technique for finding means of several variables simultaneously using a set of lower order integrals is shown to be useful for statistical estimation of Gaussian signals for interpolation. Hermite expansion based computation of probability integrals appears to be the simplest; the labor involved can be reduced by partitioning.

The work in this paper made an effort to find a way to incorporate partially known signals with various degrees of uncertainty in the interpolation process for non-uniform sampling. The key issue of finding random signal parameters requires further study and empirical confirmation.  

\section*{Acknowledgment}
The author is grateful to Mr. M. Ravi Kumar for generous editorial assistance. Thanks are also due to Mr. Arka Majumdar, Mr. Tamal Das and Mr. Lakshi Prosad Roy for help. Mr. Arka Majumdar has been associated  with the work on information of zero-crossings mainly for band pass signals; the results of the simulation of Gaussian and sub-Gaussian signals are drawn upon but are not included.

\end{document}